%% file: main.tex
\documentclass[11pt,a4paper]{article}
\usepackage[T1]{fontenc}
\usepackage{lmodern}
\usepackage[a4paper,margin=25mm,headheight=15pt]{geometry}
\usepackage{amsmath,amssymb,amsthm,bm,mathtools}
\usepackage{booktabs,tabularx,array,longtable}
\usepackage{graphicx,caption}
\usepackage{microtype,xcolor,fancyhdr,setspace}
\usepackage[round,authoryear]{natbib}
\usepackage[colorlinks=true,linkcolor=blue!40!black,citecolor=blue!40!black,urlcolor=blue!40!black]{hyperref}
\usepackage[nameinlink,noabbrev]{cleveref}
\hypersetup{
  pdftitle={Foundations of a solved-volatility stochastic turbulence closure: Ito-Hencky kinematics, source-consistent momentum and finite-correlation realisation},
  pdfauthor={Hsieh-Chen Tsai},
  pdfsubject={Theoretical foundations of a solved-volatility stochastic closure for incompressible flow},
  pdfkeywords={stochastic fluid mechanics, turbulence closure, Ito-Hencky strain, stochastic Reynolds transport, differential-algebraic equations, Green-Kubo covariance}
}

\newcommand{\dd}{\mathrm{d}}
\newcommand{\Dt}{\Delta t}
\newcommand{\bx}{\bm{x}}
\newcommand{\bX}{\bm{X}}
\newcommand{\bc}{\bm{c}}
\newcommand{\bz}{\bm{z}}
\newcommand{\bu}{\bm{u}}
\newcommand{\bv}{\bm{v}}
\newcommand{\bw}{\bm{w}}
\newcommand{\bq}{\bm{q}}
\newcommand{\bg}{\bm{g}}
\newcommand{\bn}{\bm{n}}
\newcommand{\balpha}{\bm{\alpha}}
\newcommand{\ba}{\bm{a}}
\newcommand{\bA}{\bm{A}}
\newcommand{\bL}{\bm{L}}
\newcommand{\bD}{\bm{D}}
\newcommand{\bW}{\bm{W}}
\newcommand{\bI}{\bm{I}}
\newcommand{\bT}{\bm{T}}
\newcommand{\bC}{\bm{C}}
\newcommand{\bB}{\bm{B}}
\newcommand{\bQ}{\bm{Q}}
\newcommand{\bF}{\bm{F}}
\newcommand{\bS}{\bm{S}}
\newcommand{\bSigma}{\bm{\Sigma}}
\newcommand{\bs}{\bm{s}}
\newcommand{\cT}{\bm{\mathcal T}}

\newcommand{\cH}{\mathcal H}
\newcommand{\cN}{\mathcal N}
\newcommand{\cS}{\mathcal S}
\newcommand{\cD}{\mathcal D}
\newcommand{\E}{\mathbb E}
\newcommand{\tr}{\operatorname{tr}}
\newcommand{\Tr}{\operatorname{Tr}}
\newcommand{\sym}{\operatorname{sym}}
\newcommand{\skw}{\operatorname{skw}}
\newcommand{\Div}{\operatorname{div}}
\newcommand{\inner}{\mathbin{:}}

\newtheorem{theorem}{Theorem}[section]
\newtheorem{proposition}[theorem]{Proposition}
\newtheorem{corollary}[theorem]{Corollary}
\newtheorem{lemma}[theorem]{Lemma}
\theoremstyle{definition}

\theoremstyle{remark}

\title{\bfseries Foundations of a solved-volatility stochastic turbulence closure:\\It\^o--Hencky kinematics, source-consistent momentum and finite-correlation realisation}
\author{Hsieh-Chen Tsai\\
Department of Mechanical Engineering, National Taiwan University\\
\href{mailto:hsiehchentsai@ntu.edu.tw}{hsiehchentsai@ntu.edu.tw}\\
\href{https://orcid.org/0000-0002-4240-4332}{ORCID: 0000-0002-4240-4332}}
\date{Theory preprint v1.1 --- 28 July 2026}

\begin{document}
\maketitle

\begin{abstract}
Most stochastic closures prescribe a covariance tensor, a noise basis or an eddy-viscosity field. This paper develops a different framework in which the displacement-volatility field is solved together with the resolved velocity. The starting point is a one-channel It\^o configuration map. A local matrix-logarithm expansion gives distinct material and spatial Hencky increments, their quadratic-variation drifts and the exact pathwise volume constraint. Under constant density, one Brownian channel, pathwise isochoricity and no independent martingale in the resolved Eulerian drift, the material pull-back momentum equation is shown to be the on-shell form of the full stochastic Reynolds transport balance when the momentum-source covariation is retained. The resulting velocity--volatility--pressure system has an index-one differential--algebraic structure. Virtual power fixes the mechanical type of the stress impulse and separates work from quadratic covariation. A finite-correlation precursor then gives a Green--Kubo realisation of the solved displacement covariance. State dependence adds a Lyapunov noise-induced drift, while dynamic boundaries add reaction work and active/passive covariance compatibility conditions. The analysis also gives four limits: a non-zero Brownian transport limit cannot retain finite ordinary unresolved kinetic energy; total energy alone does not fix entropy production; wall tangency limits covariance rank rather than the number of stochastic modes; and a homogeneous decoupled Helmholtz--Stokes equation has only the trivial periodic solution. The result is a theory-complete, testable closure architecture. Developed turbulent statistics, logarithmic wall scaling and computational-fluid-dynamics validation are deliberately left to the expanded fluid-mechanics study.
\end{abstract}

\section*{Nomenclature}
\small
The table is a reference only. Every abbreviation is also written in full at first use. Bold lower-case symbols denote vectors and bold upper-case symbols denote second-order tensors or finite-dimensional operators unless stated otherwise.
\begin{longtable}{@{}>{\raggedright\arraybackslash}p{0.22\linewidth}p{0.72\linewidth}@{}}
$t$, $\bx_t$ & time and stochastic particle position \\
$\bu$, $\balpha$ & resolved It\^o displacement drift and solved displacement-volatility field \\
$W_t$, $\bm W_t$ & scalar and vector or cylindrical standard Brownian motions \\
$\ba=\balpha\otimes\balpha$ & one-channel displacement covariance per unit time \\
$\bL=\nabla\bu$, $\bA=\nabla\balpha$ & velocity-gradient and volatility-gradient tensors \\
$\bD_u=\sym\bL$, $\bD_\alpha=\sym\bA$, $\bW_\alpha=\skw\bA$ & resolved stretching, volatility stretching and volatility spin \\
$\bF$, $J=\det\bF$ & local relative deformation gradient and local volume ratio \\
$\bm H$, $\bm h$ & local material and spatial Hencky increments \\
$\bs=\tfrac12\nabla\cdot\ba$, $\bv=\bu-\bs$ & It\^o--Stokes correction and divergence-free transport velocity \\
$\dd\cT=\bT_0\,\dd t+\bT_1\,\dd W_t$ & stress--time semimartingale and its drift and martingale coefficients \\
$p_0$, $p_1$ & drift and martingale pressure channels \\
$\rho$, $\mu$, $\nu$ & density, dynamic viscosity and kinematic viscosity \\
$\bQ_0$, $\bQ_1$ & drift and martingale momentum-source coefficients \\
$R$, $K$, $C$, $C^*$ & resolved damping, fast relaxation, resolved--unresolved coupling and its adjoint \\
$G$, $A$, $\bSigma_q$ & positive relaxation operator, target Green--Kubo covariance and stationary fast covariance \\
$\tau_c$, $\bq^\tau$ & correlation time and finite-correlation unresolved velocity \\
$\cS$, $\cD$, $J_f$, $\cN$ & feedback-corrected fast operator, slow response, frozen covariance and noise-induced drift \\
$N$, $L$, $T=L^*$ & homogeneous basis map, wall lifting and wall trace \\
$\bg$, $\bm\Lambda_\Gamma$ & wall state and wall reaction impulse \\
$M_\Gamma$, $Z_\Gamma$, $B_\Gamma$ & wall mass, impedance and noise amplitude \\
$A_\Gamma^{\rm disp}$, $A_\Gamma^{\rm pass}$ & active target and attainable passive wall displacement covariances \\
\addlinespace[4pt]
\multicolumn{2}{@{}l}{\textit{Abbreviations}}\\[2pt]
DAE & differential--algebraic equation \\
LU & modelling under location uncertainty \\
OU & Ornstein--Uhlenbeck \\
SALT & stochastic advection by Lie transport \\
SRTT & stochastic Reynolds transport theorem \\
\end{longtable}
\normalsize

\input{theory_core.tex}

\section*{Acknowledgements}
The author used generative artificial-intelligence tools for language editing, document organisation, symbolic checking and code assistance. The author determined the scientific scope, checked the derivations and accepts responsibility for the manuscript.

\section*{Funding}
This research received no specific grant from any funding agency in the public, commercial or not-for-profit sectors.

\section*{Declaration of interests}
The author reports no conflict of interest.

\section*{Data and code availability}
The source package contains the LaTeX manuscript, bibliography and figures needed to reproduce this preprint. Symbolic and finite-dimensional verification code will be linked to the arXiv record when the public repository is released. No computational-fluid-dynamics dataset is claimed in this theory version.

\bibliographystyle{plainnat}
\bibliography{references}
\end{document}

%% file: theory_core.tex
\section{Introduction}
\label{sec:introduction}

The turbulence closure problem is usually stated after averaging or filtering the Navier--Stokes equations. The nonlinear momentum flux then introduces a Reynolds stress or subgrid stress that is not known from the resolved variables alone. Common closures prescribe an eddy viscosity, selected moment equations or a filter-dependent stress model \citep{Pope2000,Smagorinsky1963,GermanoEtAl1991,MeneveauKatz2000}. These methods are useful, but the unresolved transport remains a modelling input.

Stochastic transport starts from a different description. In modelling under location uncertainty (LU), the particle displacement is separated into a smooth drift and a temporally decorrelated spatial field. The stochastic Reynolds transport theorem (SRTT) then gives conservative balance laws with covariance transport and an It\^o--Stokes correction \citep{Memin2014,ResseguierEtAl2017a,TissotEtAl2024,TissotEtAl2026}. Stochastic advection by Lie transport (SALT) provides a variational route in which stochastic vector fields transport momentum one-forms and preserve selected geometric structures \citep{Holm2015,StreetCrisan2021,DrivasHolm2020}. In both families, the spatial noise is commonly prescribed, estimated or assigned a separate evolution law.

This paper studies the alternative in which the displacement volatility itself is a continuum unknown. The covariance is derived only after the volatility has been solved. This distinction matters because the local gradient of each volatility mode enters deformation, traction and source--transport covariation. A covariance tensor alone does not retain this information.

\textbf{The central modelling statement is simple: displacement volatility is solved with the flow rather than prescribed as turbulence data.} The one-channel configuration map is
\begin{equation}
  \dd\bx_t=\bu(\bx_t,t)\,\dd t+\balpha(\bx_t,t)\,\dd W_t,
  \label{eq:particle-map}
\end{equation}
where $\bu$ is the It\^o displacement drift and $\balpha$ is the solved volatility. The covariance $\ba=\balpha\otimes\balpha$ is rank one at each point. Rank one is a minimal model choice, not a universal claim. It keeps the theory falsifiable while allowing $\nabla\balpha$ to contain general stretching, shear and rotation.

The term \emph{closure} is used in a restricted field-theoretic sense. Molecular properties, geometry, body forces, initial data and boundary receptivity remain physical inputs. The unknowns are the resolved velocity, displacement volatility and two pressure channels. No interior eddy-viscosity profile, Reynolds stress, covariance profile, mixing length or von K\'arm\'an constant is supplied. A square field count does not prove uniqueness, developed turbulence, a logarithmic law or a complete entropy theorem.

The contributions are as follows. First, a local matrix-logarithm calculation gives material and spatial It\^o--Hencky increments and the exact pathwise Jacobian constraint. Second, virtual power and source consistency fix the mechanical role of the stress impulse and the momentum-source covariation. Third, the resulting field equations have an index-one differential--algebraic equation (DAE) form. Fourth, a finite-correlation construction supplies a Green--Kubo interpretation of the displacement covariance, including the corrections required by state dependence and dynamic boundaries. Finally, several no-go results state what the framework does not provide automatically.

\begin{figure}[t]
  \centering
  \includegraphics[width=0.96\linewidth]{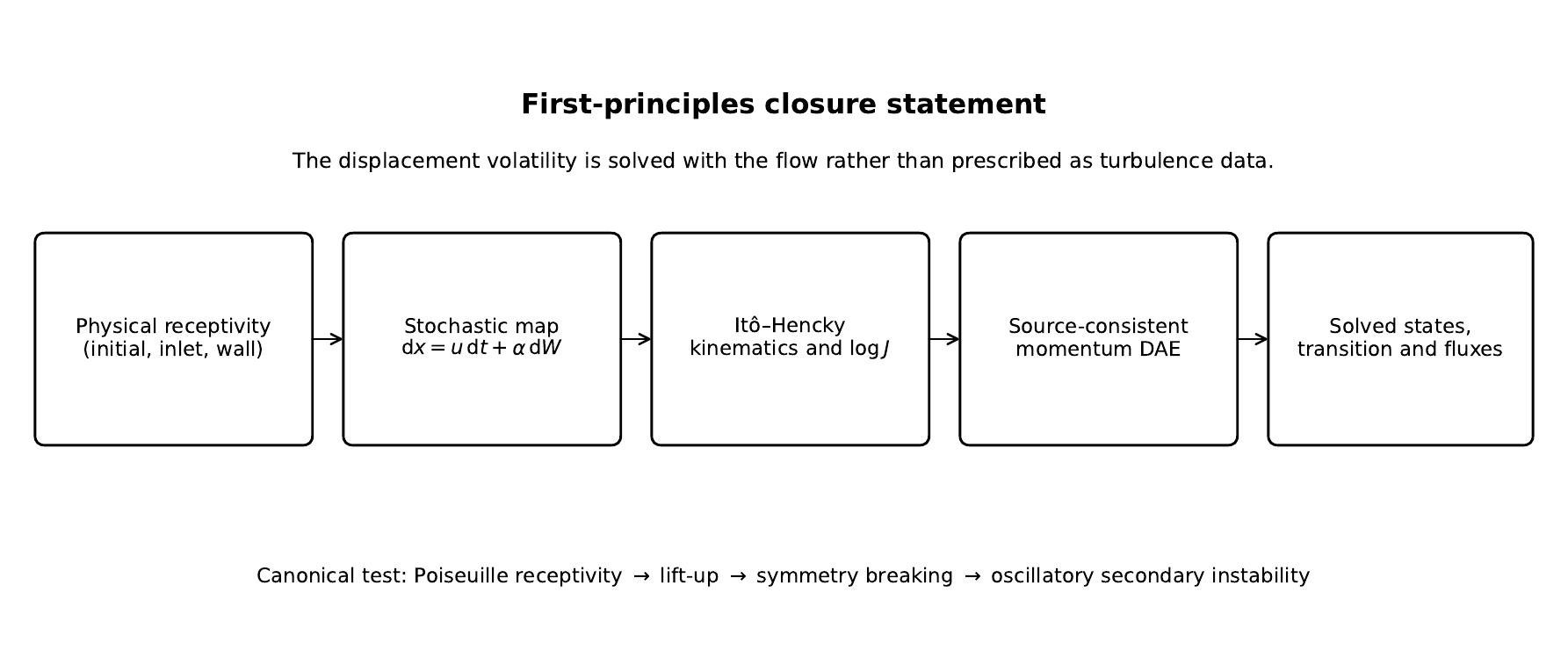}
  \caption{Closure architecture. The present paper establishes the stochastic map, It\^o--Hencky kinematics, source-consistent momentum and finite-correlation interpretation. The later full fluid-mechanics version adds the physical channel-flow calculations represented by the final box.}
  \label{fig:architecture}
\end{figure}

\section{Local It\^o--Hencky kinematics}
\label{sec:kinematics}

Set $\bL=\nabla\bu$, $\bA=\nabla\balpha$, $\bD_u=\sym\bL$, $\bD_\alpha=\sym\bA$ and $\bW_\alpha=\skw\bA$. Over a local interval, the relative deformation is
\[
  \bF=\bI+\bA\,\dd W_t+\bL\,\dd t.
\]
The right and left Cauchy--Green tensors differ at order $\dd t$ because the It\^o quadratic variation keeps $\bA^{\mathsf T}\bA$ or $\bA\bA^{\mathsf T}$ according to the ordering.

\begin{theorem}[Local material and spatial Hencky increments]
\label{thm:hencky}
For sufficiently smooth $\bu$ and $\balpha$, and before the local deformation loses orientation, the material and spatial Hencky increments through order $\dd t$ are
\begin{equation}
\begin{aligned}
  \dd\bm H
  &=\bD_\alpha\,\dd W_t+
  \left(\bD_u+\tfrac12\bA^{\mathsf T}\bA-\bD_\alpha^2\right)\dd t,\\
  \dd\bm h
  &=\bD_\alpha\,\dd W_t+
  \left(\bD_u+\tfrac12\bA\bA^{\mathsf T}-\bD_\alpha^2\right)\dd t.
\end{aligned}
\label{eq:hencky}
\end{equation}
Denote the quadratic drift corrections in the two lines by $\bQ_H$ and $\bQ_h$, respectively. Their difference is
\begin{equation}
  \bQ_h-\bQ_H
  =\tfrac12(\bA\bA^{\mathsf T}-\bA^{\mathsf T}\bA)
  =\bW_\alpha\bD_\alpha-\bD_\alpha\bW_\alpha.
  \label{eq:left-right}
\end{equation}
The difference is symmetric and trace free, and it vanishes when $\bA$ is normal.
\end{theorem}

The proof is given in \cref{app:matrixlog}. The result is local. Ordinary polar rotation does not have a dynamic superposition property, so \cref{eq:hencky} is not by itself a finite-time non-coaxial history law \citep{Haller2016}.

\begin{corollary}[Logarithmic Jacobian and pathwise volume preservation]
\label{cor:jacobian}
The common trace of the two increments gives
\begin{equation}
  \dd\log J=(\nabla\cdot\balpha)\,\dd W_t+
  \left[\nabla\cdot\bu-\tfrac12\tr(\bA^2)\right]\dd t.
  \label{eq:logJ}
\end{equation}
Hence pathwise isochoricity is equivalent to
\[
  \nabla\cdot\balpha=0,
  \qquad
  \nabla\cdot\bu=\tfrac12\tr(\bA^2).
\]
If $\bs=\tfrac12\nabla\cdot(\balpha\otimes\balpha)$ and $\bv=\bu-\bs$, the same conditions give $\nabla\cdot\bv=0$.
\end{corollary}

The distinction between $\bu$ and $\bv$ is essential. The It\^o displacement drift need not be divergence free even though the stochastic flow preserves volume. Imposing $\nabla\cdot\bu=0$ as an additional constraint would generally conflict with \cref{eq:logJ}.

\subsection{Observer covariance}

Let $\bx^*=\bm Q(t)\bx+\bc(t)$, where $\bm Q\bm Q^{\mathsf T}=\bI$, and define $\bm\Omega=\dot{\bm Q}\bm Q^{\mathsf T}$. Then
\[
  \bu^*=\bm Q\bu+\bm\Omega(\bx^*-\bc)+\dot\bc,
  \qquad
  \balpha^*=\bm Q\balpha.
\]
Because $\bm\Omega$ is skew, it drops from the stretching part of $\nabla^*\bu^*$.

\begin{proposition}[Coefficient-level objectivity]
\label{prop:objectivity}
The martingale and drift coefficients in \cref{eq:hencky} transform by orthogonal conjugation. In particular,
\begin{equation}
  (\dd\bm h)^*=\bm Q\,\dd\bm h\,\bm Q^{\mathsf T},
  \qquad
  (\dd\bm H)^*=\bm Q\,\dd\bm H\,\bm Q^{\mathsf T}
  \label{eq:objectivity}
\end{equation}
when the material tensor is expressed in the rotated material basis.
\end{proposition}

This proposition concerns local constitutive tensors. Full covariance of a non-inertial momentum equation also requires the corresponding inertial body forces \citep{Murdoch2003,TruesdellNoll2004}.

\subsection{Logarithmic work-conjugate differential}

In deterministic finite-strain mechanics, the Eulerian Hencky strain is work conjugate to the Cauchy stress only through the logarithmic corotational rate \citep{XiaoEtAl1997,Norris2008,BruhnsXiaoMeyers2002}. A canonical one-channel smooth-noise limit gives a neighbouring stochastic differential
\[
  \mathcal D^{\log}\bm h
  =\bD_u\,\dd t+\bD_\alpha\circ\dd W_t.
\]
In It\^o form its quadratic drift is
\begin{equation}
  \bQ_{\log}=-\tfrac12\sym(\bA^2)
  =-\tfrac12(\bD_\alpha^2+\bW_\alpha^2)
  =\tfrac12(\bQ_h+\bQ_H).
  \label{eq:qlog}
\end{equation}
Thus the work-conjugate drift lies midway between the raw material and spatial drifts. This construction is conditional on the stated one-channel Wong--Zakai limit; multi-channel rough noise can retain L\'evy-area corrections \citep{KellyMelbourne2016,DiamantakisWoodfield2025}.

\section{Stress impulse, virtual power and constitutive routes}
\label{sec:virtualpower}

Write the stress-time measure as $\dd\cT=\bT_0\,\dd t+\bT_1\,\dd W_t$. For a predictable virtual velocity $\bw$ on a fixed domain $\Omega$, the divergence theorem gives
\begin{equation}
  \int_{\partial\Omega}\bw\cdot(\dd\cT\,\bn)\,\dd a
  =\int_\Omega\left[\bw\cdot\Div\dd\cT+\nabla\bw\inner\dd\cT\right]\dd V.
  \label{eq:virtualpower}
\end{equation}
If the stress coefficients are symmetric, the internal term is $\bD(\bw)\inner\dd\cT$, where $\bD(\bw)=\sym\nabla\bw$ \citep{Lidstrom2012}.

\begin{proposition}[Mechanical type of the stress impulse]
\label{prop:worktype}
A stress-time measure acts on a velocity or strain rate. If $\dd\bm h=\bm h_0\,\dd t+\bm h_1\,\dd W_t$, then
\[
  \dd\cT\inner\dd\bm h=(\bT_1\inner\bm h_1)\,\dd t.
\]
This quantity is the quadratic covariation $\dd[\cT,\bm h]_t$, has units of stress times time and vanishes in the deterministic limit. It is not the mechanical-work increment.
\end{proposition}

Two neighbouring constitutive routes can therefore be stated without confusing work and covariation. The raw spatial It\^o--Hencky route is
\begin{equation}
\begin{aligned}
  \bT_1&=-p_1\bI+2\mu\bD_\alpha,\\
  \bT_0^{(h)}&=-p_0\bI+2\mu(\bD_u+\bQ_h),
\end{aligned}
\label{eq:route-h}
\end{equation}
whereas the logarithmic-corotational route replaces $\bQ_h$ by $\bQ_{\log}$. Both are symmetric, objective and Newtonian when $\balpha=0$. Objectivity and the deterministic limit do not select between them. The expanded fluid-mechanics study treats this as a robustness question rather than claiming that stochastic transport alone fixes the constitutive law.

\section{Source-consistent momentum and the index-one field system}
\label{sec:momentum}

Assume constant density, one scalar Brownian channel, pathwise isochoricity and no independent martingale component in the resolved Eulerian drift. Let $\bQ_0\,\dd t+\bQ_1\,\dd W_t$ be the momentum source, and let $\bm b_0$ and $\bm b_1$ be prescribed body-force coefficients per unit mass. The material pull-back coefficients are
\begin{equation}
\begin{aligned}
 \rho\left[\partial_t\bu+(\bu\cdot\nabla)\bu+
 \tfrac12(\balpha\otimes\balpha)\inner\nabla\nabla\bu\right]
 &=\Div\bT_0+\rho\bm b_0,\\
 \rho(\balpha\cdot\nabla)\bu
 &=\Div\bT_1+\rho\bm b_1.
\end{aligned}
\label{eq:pullback}
\end{equation}

The full SRTT includes the martingale source and the source--transport quadratic covariation. Its martingale coefficient requires
\[
  \bQ_1=\rho(\balpha\cdot\nabla)\bu.
\]
Under $\nabla\cdot\balpha=0$,
\[
  \rho^{-1}\Div(\bQ_1\otimes\balpha)=\Div[(\balpha\otimes\balpha)\nabla\bu].
\]
This term restores the covariance flux that is lost if the source and transport are separated before taking quadratic variation.

\begin{theorem}[Restricted source-consistency equivalence]
\label{thm:source}
Under the assumptions stated above, the material pull-back balance \cref{eq:pullback} and the full source-consistent SRTT are equivalent on shell. A reduced LU equation that omits, parameterises or reallocates $\Div(\bQ_1\otimes\balpha)$ is a different closure convention.
\end{theorem}

Combining \cref{eq:pullback} with either constitutive route and the two volume constraints gives a closed field system for $(\bu,\balpha,p_0,p_1)$. With isotropic quadratic terms absorbed into a modified drift pressure $\pi_0$, and with $\bT_\alpha=\bA\bA^{\mathsf T}-2\bD_\alpha^2$, the raw spatial route can be written
\begin{equation}
\begin{aligned}
 \rho\left[\partial_t\bu+(\bu\cdot\nabla)\bu+
 \tfrac12\ba\inner\nabla\nabla\bu\right]
 &=-\nabla\pi_0+\mu\Delta\bu+\mu\Div\bT_\alpha+\rho\bm b_0,\\
 \rho(\balpha\cdot\nabla)\bu
 &=-\nabla p_1+\mu\Delta\balpha+\rho\bm b_1,\\
 \nabla\cdot\balpha&=0,
 \qquad
 \nabla\cdot\bu=\tfrac12\tr[(\nabla\balpha)^2],
\end{aligned}
\label{eq:closed-system}
\end{equation}
\subsection{Differential--algebraic structure}

The martingale momentum equation, the martingale pressure and both divergence constraints are instantaneous compatibility conditions. After spatial discretisation, the Newton system has differential state $\bm y$, algebraic state $\bm z$, residuals $\bm F_d$ and $\bm F_a$, and differential--algebraic blocks,
\[
  \begin{pmatrix}
  \bA_{\Dt} & \bB\\
  \bC & \bD
  \end{pmatrix}
  \begin{pmatrix}\delta\bm y\\\delta\bm z\end{pmatrix}
  =-\begin{pmatrix}\bm F_d\\\bm F_a\end{pmatrix}.
\]
If $\bD^\dagger$ is the constrained minimum-norm inverse of the algebraic Stokes block, the differential correction uses
\begin{equation}
  \bS_{\Dt}=\bA_{\Dt}-\bB\bD^\dagger\bC.
  \label{eq:schur}
\end{equation}
Exact pressure and potential gauges remain in the null space. When the algebraic Jacobian has constant rank modulo those gauges, the system is index one. This structure is important for numerical continuation, but it is also part of the closure: $\balpha$ is not advanced by an independent diffusion equation and then inserted into momentum. It is selected on the coupled algebraic manifold.

\section{Resolved energy and the thermodynamic boundary}
\label{sec:energy}

Let the resolved momentum source be $\bQ_0\,\dd t+\bQ_1\,\dd W_t$, and write the resolved kinetic energy as $\mathcal K_u$. It\^o's formula gives on a material volume,
\begin{equation}
\begin{aligned}
 \dd\mathcal K_u
 ={}&\left[\int\left(\bu\cdot\bQ_0+\frac{|\bQ_1|^2}{2\rho}\right)\dd V\right]\dd t
 +\left[\int\bu\cdot\bQ_1\,\dd V\right]\dd W_t.
\end{aligned}
\label{eq:resolved-energy}
\end{equation}
If $\bQ_r=\Div\bT_r+\rho\bm b_r$ for $r=0,1$, integration by parts produces stress work, body-force work and boundary traction work in both stochastic channels. The quadratic term $|\bQ_1|^2/(2\rho)$ is mandatory. It is a resolved energy input and not automatically a total dissipation.

In the pathwise-isochoric setting,
\[
  p_0\nabla\cdot\bu+\bs\cdot\nabla p_0=\nabla\cdot(p_0\bs).
\]
Thus the apparent pressure conversion and It\^o--Stokes alignment combine into a boundary flux in a periodic or compatible no-flux domain. The identity does not extend unchanged to compressible flow or arbitrary open boundaries.

\begin{proposition}[Thermodynamic reservoir non-uniqueness]
\label{prop:reservoir}
Suppose internal energy $\mathcal U$ and unresolved energy $\mathcal E_\alpha$ close a total-energy balance. For any adapted semimartingale $\chi$, the replacement
\[
  \dd\mathcal U\mapsto\dd\mathcal U+\dd\chi,
  \qquad
  \dd\mathcal E_\alpha\mapsto\dd\mathcal E_\alpha-\dd\chi
\]
leaves the total energy unchanged. The first law alone therefore does not fix the internal state or entropy production.
\end{proposition}

A Gibbs relation, heat flux, boundary covariations and a fluctuation--dissipation law are still required for a fluid Clausius--Duhem theorem \citep{GrmelaOttinger1997,Ottinger1998,SerranoEspanol2001}.

\section{Finite-correlation realisation}
\label{sec:finite}

\subsection{Scalar transport limit and energy singularity}

A scalar Ornstein--Uhlenbeck (OU) precursor is
\begin{equation}
  \dd\xi_t^\tau=-\tau_c^{-1}\xi_t^\tau\,\dd t+\tau_c^{-1}\dd W_t,
  \qquad
  \bq^\tau=\balpha\xi_t^\tau.
  \label{eq:scalar-ou}
\end{equation}
It satisfies
\[
  \int_0^t\xi_s^\tau\,\dd s=W_t-\tau_c(\xi_t^\tau-\xi_0^\tau),
\]
so the integrated coloured velocity converges in mean square to $\balpha W_t$. At stationarity,
\[
  \E(\bq^\tau\otimes\bq^\tau)=\frac{\balpha\otimes\balpha}{2\tau_c},
  \qquad
  \E\frac{\rho|\bq^\tau|^2}{2}=\frac{\rho|\balpha|^2}{4\tau_c}.
\]

\begin{theorem}[Transport--energy dichotomy]
\label{thm:transport-energy}
A non-zero Brownian displacement limit and finite ordinary instantaneous unresolved kinetic energy cannot both be retained as $\tau_c\to0$. The white-noise model is a displacement-transport limit, not a finite-energy instantaneous-velocity limit.
\end{theorem}

The OU density $f$ has stationary Gaussian $f_\infty$, and its relative entropy satisfies
\begin{equation}
  \frac{\dd}{\dd t}\cH(f\mid f_\infty)
  =-\frac{1}{2\tau_c^2}\int f\left|\partial_\xi\log\frac{f}{f_\infty}\right|^2\dd\xi\le0.
  \label{eq:ou-entropy}
\end{equation}
This is an entropy result for the auxiliary stochastic sector, not a full fluid entropy theorem.

\begin{figure}[t]
  \centering
  \includegraphics[width=0.70\linewidth]{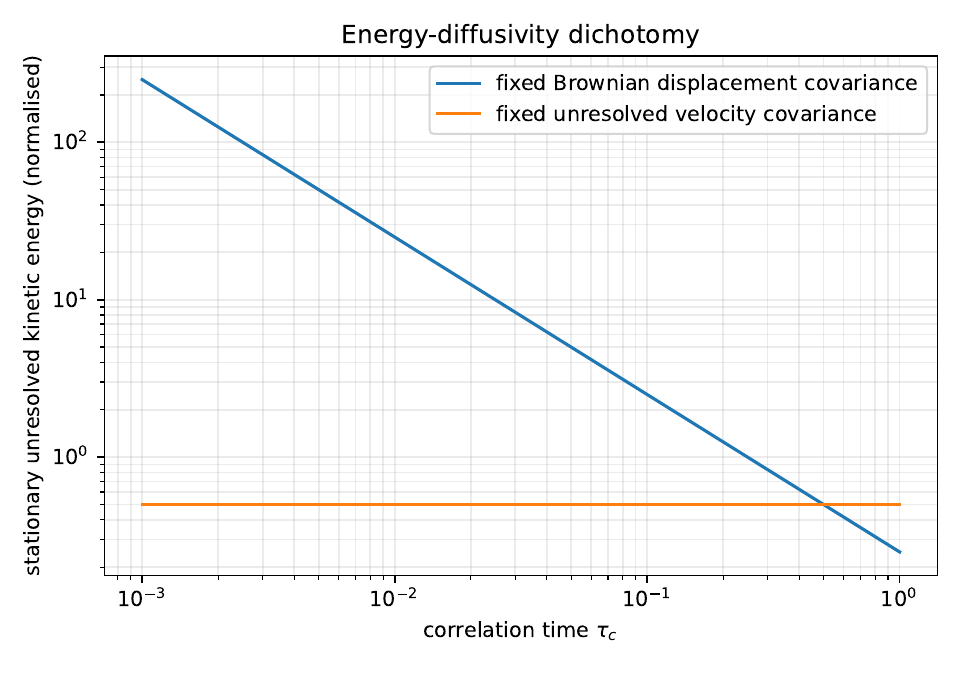}
  \caption{Finite-correlation scaling. The integrated displacement converges to the Brownian target, while the ordinary unresolved kinetic energy grows as the correlation time decreases.}
  \label{fig:fc-scaling}
\end{figure}

\subsection{Operator Green--Kubo calibration}

Let $G$ be positive and self-adjoint, and let $A$ be the target two-sided displacement covariance. Consider
\begin{equation}
  \dd\bq=-G\bq\,\dd t+GA^{1/2}\dd\bm W_t.
  \label{eq:operator-ou}
\end{equation}
The stationary covariance $\bSigma_q$ solves
\[
  G\bSigma_q+\bSigma_qG=GAG.
\]

\begin{theorem}[Green--Kubo realisation]
\label{thm:gk}
If the Lyapunov equation has a trace-class stationary solution, the two-sided Green--Kubo covariance of \cref{eq:operator-ou} is exactly
\begin{equation}
  \int_{-\infty}^{\infty}\E[\bq(t)\otimes\bq(0)]\,\dd t
  =G^{-1}\bSigma_q+\bSigma_qG^{-1}=A.
  \label{eq:gk}
\end{equation}
\end{theorem}

For the one-channel target $A=\balpha\otimes\balpha$, the covariance is finite rank. If $\balpha\in D(G)$, the noise covariance has finite trace. An equal-temperature ultraviolet divergence found by independently forcing every Fourier degree of freedom is therefore not intrinsic to the one-channel model.

For an isotropic multi-channel spectrum $A_s\sim(\ell^{-2}I-\Delta)^{-s}$ and relaxation $G_r\sim(\ell_g^{-2}I-\Delta)^r$ in $d$ dimensions, finite injection requires
\begin{equation}
  s>2r+\frac d2.
  \label{eq:trace-class}
\end{equation}
In three dimensions with Laplacian-order relaxation, this gives $s>7/2$. A fractional operator may act as a covariance precision, but equipartition in that metric does not remove mechanical-energy divergence.

\begin{figure}[t]
  \centering
  \includegraphics[width=0.70\linewidth]{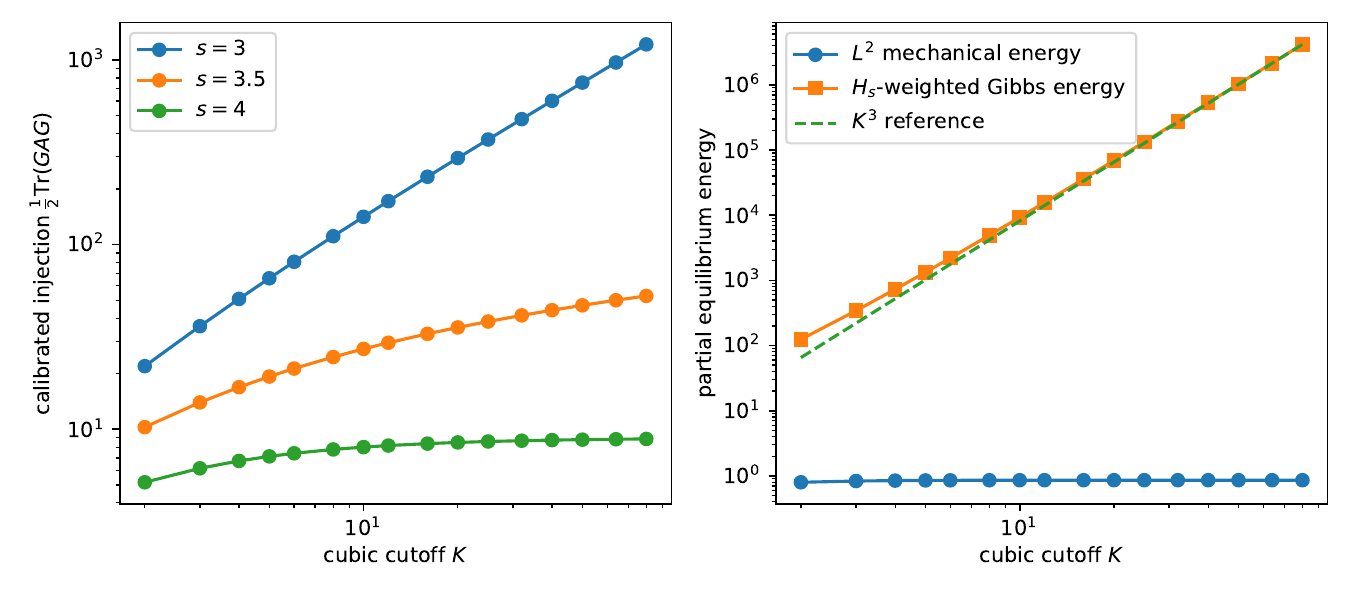}
  \caption{Trace-class gate for an isotropic multi-channel extension. Finite injection in three dimensions with Laplacian-order relaxation requires a spectrum steeper than $s=7/2$.}
  \label{fig:trace-class}
\end{figure}

\subsection{Resolved feedback and state dependence}

In a linear coupled block,
\[
  \dd\bw=(-R\bw+C\bq)\,\dd t,
  \qquad
  \dd\bq=(-C^*\bw-K\bq)\,\dd t+B\,\dd W_t.
\]
Resolved feedback changes the zero-frequency calibration.

\begin{proposition}[Feedback-corrected calibration]
\label{prop:feedback}
If $R$ is invertible on the constrained resolved subspace and the coupled drift is stable, the noise amplitude
\begin{equation}
  B=(K+C^*R^{-1}C)\balpha
  \label{eq:feedback}
\end{equation}
recovers the target displacement covariance $\balpha\otimes\balpha$. The adjoint pair $C$ and $-C^*$ makes the resolved--unresolved mechanical exchange cancel exactly.
\end{proposition}

If $R$, $K$, $C$ and $\balpha$ depend on a slow stochastic state $\bm\zeta$, pointwise use of \cref{eq:feedback} is not enough. Set $\cS=K+C^*R^{-1}C$, $B=\cS\balpha$ and $\cD=C\cS^{-1}$. Let $J_f$ solve the frozen Lyapunov equation $\cS J_f+J_f\cS^*=BB^*$.

\begin{theorem}[State-dependent noise-induced drift]
\label{thm:noise-drift}
Under the standard fast--slow regularity and stability assumptions, adiabatic elimination adds the drift
\begin{equation}
  \cN_i=\partial_{\zeta_\ell}\cD_{ij}\,C_{\ell k}(J_f)_{kj}.
  \label{eq:noise-drift}
\end{equation}
A coloured precursor converges to the intended It\^o equation only after $-\cN$ is included in the prelimit slow drift. If the constraint projector moves with the state, the derivative must be covariant and the It\^o connection drift must also be retained.
\end{theorem}

This result follows the general structure of noise-induced drift in fast--slow stochastic systems \citep{HottovyEtAl2015,BirrellEtAl2017}. It also explains why a frozen Green--Kubo match is not a complete nonlinear closure.

\section{Dynamic boundary port}
\label{sec:boundary}

Let $N$ map homogeneous unresolved coordinates $\bz$ into the bulk, let $L$ lift the wall state $\bg$, and let $T=L^*$ be the wall trace. Thus $\bq=N\bz+L\bg$, with $TL=I$ and $TN=0$. Let $\dd\bm\Lambda_\Gamma$ be the wall reaction impulse. At fixed geometry,
\begin{equation}
\begin{aligned}
 \dd\bq&=\bm f_q\,\dd t+L\,\dd\bm\Lambda_\Gamma,
 &T\bq&=\bg,\\
 M_\Gamma\dd\bg&=-Z_\Gamma\bg\,\dd t+B_\Gamma\dd\bm W_t-\dd\bm\Lambda_\Gamma.
\end{aligned}
\label{eq:wall-port}
\end{equation}

\begin{proposition}[Reaction work]
\label{prop:reaction}
The reaction enters the fluid and wall with opposite signs. Its total power is
\[
  (T\bq-\bg)\cdot\dd\bm\Lambda_\Gamma=0.
\]
The physical boundary port is $\bg\cdot\dd\bm\Lambda_\Gamma$. The lifting term $\langle\bq,L\,\dd\bg\rangle$ is a coordinate-inertia term and is not generally equal to traction work.
\end{proposition}

After the reaction is eliminated, let $\mathsf A_\tau$ be the stable drift, $\mathsf N_\Gamma$ the boundary input and $E_\Gamma$ the wall output. The zero-frequency transfer is $H_\Gamma(0)=E_\Gamma(-\mathsf A_\tau)^{-1}\mathsf N_\Gamma$. On its reachable tangent range, an active boundary uses
\begin{equation}
  B_\Gamma=H_\Gamma(0)^\dagger(A_\Gamma^{\rm disp})^{1/2}.
  \label{eq:active-wall}
\end{equation}
This matches a prescribed displacement covariance but represents an externally powered actuator.

For one passive tangent channel with fluid impedance $Z_f$, wall impedance $Z_\Gamma$ and fluctuation--dissipation amplitude $B_\Gamma^2=2\vartheta Z_\Gamma$, the attainable covariance is
\begin{equation}
  A_\Gamma^{\rm pass}(Z_\Gamma)=
  \frac{2\vartheta Z_\Gamma}{(Z_f+Z_\Gamma)^2}
  \le\frac{\vartheta}{2Z_f}.
  \label{eq:passive-wall}
\end{equation}
Targets below the maximum have two positive impedance branches, the maximum has one, and larger targets have none. Active and passive boundaries are different physical models and cannot be forced to share the same covariance without changing their power law.

\begin{figure}[t]
  \centering
  \includegraphics[width=0.72\linewidth]{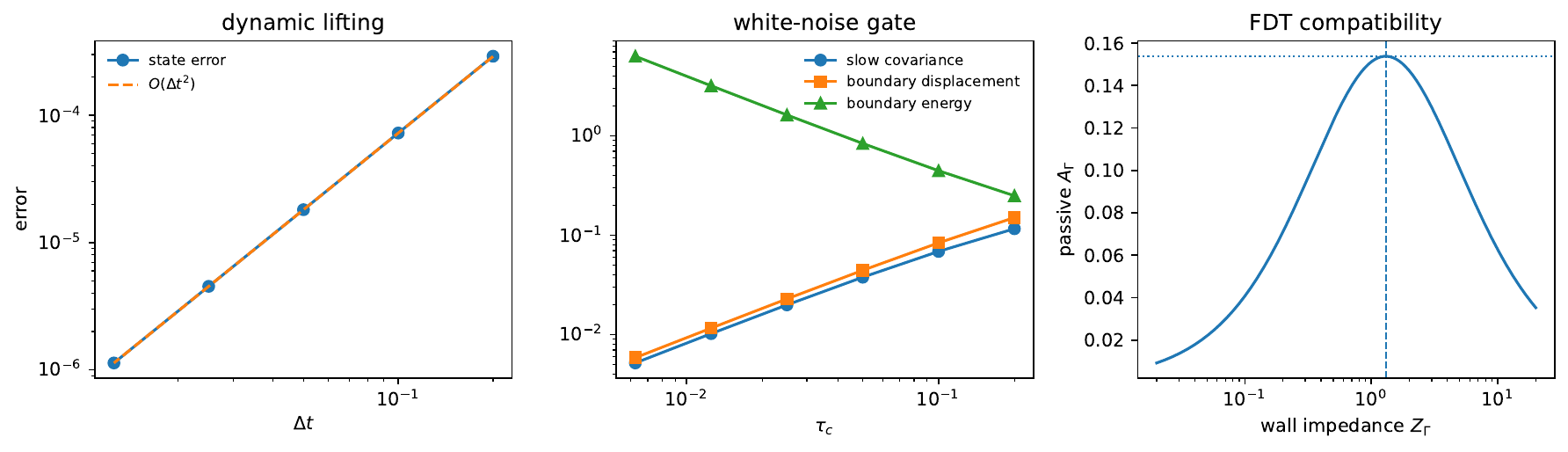}
  \caption{Dynamic-boundary gate. The complete traction port gives the energy-consistent result; the white-noise displacement limit is regular while boundary kinetic energy is singular; and a passive wall has a bounded, non-monotone covariance response.}
  \label{fig:boundary-gate}
\end{figure}

\section{Structural no-go results and scope}
\label{sec:nogo}

The following results prevent several tempting but unsupported extensions.

\begin{lemma}[Boundary covariance rank]
\label{lem:boundary-rank}
For $m$ stochastic channels, if every mode is tangent to a smooth boundary, $\balpha_r\cdot\bn=0$, then
\[
  \ba\bn=0,
  \qquad
  \operatorname{rank}(\ba|_{\partial\Omega})\le d-1,
  \qquad
  \ba=\sum_{r=1}^m\balpha_r\otimes\balpha_r.
\]
The condition limits the covariance range, not the number $m$ of Brownian modes.
\end{lemma}

\begin{lemma}[Homogeneous elliptic closure is trivial]
\label{lem:elliptic}
On a periodic domain, let $\tau_e>0$ and suppose
\[
  \nu\Delta\balpha-\tau_e^{-1}\balpha=\nabla\pi,
  \qquad
  \nabla\cdot\balpha=0.
\]
Then $\balpha=0$ and $\nabla\pi=0$.
\end{lemma}

The proof is the $L^2$ energy identity. A forcing-free homogeneous Helmholtz--Stokes block cannot create turbulence or a non-zero volatility state. Non-zero solutions require forcing, non-homogeneous boundary data or coupling to the resolved field.

These facts also set the publication boundary of the present paper. The theory does not derive a universal logarithmic law, a von K\'arm\'an constant, the absence of two-dimensional turbulence or a $k^{-5/3}$ spectrum. Such statements require physical solutions, scale separation, energy-flux diagnostics and Reynolds-number tests. They cannot be obtained from wall topology or from the local Lie algebra alone.

\section{Discussion and relation to the expanded fluid-mechanics version}
\label{sec:discussion}

The solved-volatility framework has a clear theoretical chain. The stochastic configuration map produces local material and spatial deformation measures. Their common trace gives pathwise incompressibility. Source consistency ties the martingale momentum equation to its transport covariation and yields an index-one field system. Virtual power fixes the mechanical role of the stress impulse. A finite-correlation precursor then separates displacement covariance from instantaneous velocity energy, while state dependence and boundaries add the drift and reaction terms needed for a consistent limit.

The theory is intentionally not presented as a completed turbulence prediction. The constitutive choice between the raw spatial and logarithmic-corotational drifts remains a physical robustness question. The white-noise model is a transport limit and does not have finite ordinary unresolved kinetic energy. A total-energy law does not by itself determine an entropy law. Boundary covariance also depends on whether the wall is an active actuator or a passive material.

This arXiv version is the theory-complete first stage of one manuscript lineage. The expanded Journal of Fluid Mechanics version will retain the theory developed here and add the full channel-flow formulation, computational fluid dynamics, nonlinear branch continuation, secondary instability, long-time post-secondary statistics and high-friction-Reynolds-number wall-transport tests. The two documents are therefore not separate competing papers: the journal article is intended to be the complete theoretical and physical version.

\section{Conclusions}

A stochastic closure can be organised around a solved displacement-volatility field rather than a prescribed covariance. The local It\^o--Hencky expansion supplies the quadratic deformation terms and exact pathwise volume constraint. Under stated restrictions, the material pull-back and full source-consistent SRTT are the same on-shell momentum balance, and the resulting field problem has an index-one DAE structure.

The stress-time measure must be paired with a rate; multiplying it by another stochastic increment gives quadratic covariation rather than work. A finite-correlation unresolved velocity can reproduce the solved displacement covariance through Green--Kubo calibration, but its ordinary kinetic energy is singular in the zero-correlation-time limit. State-dependent calibration requires a Lyapunov noise-induced drift, and dynamic boundaries require reaction work and an explicit active or passive constitutive law.

These results define the mathematical architecture and its limits. They establish the theoretical priority of the solved-volatility approach while leaving the decisive turbulence claims to direct fluid-mechanical validation.

\appendix
\section{Matrix-logarithm derivation}
\label{app:matrixlog}

The right Cauchy--Green tensor is
\[
  \bF^{\mathsf T}\bF
  =\bI+2\bD_\alpha\,\dd W_t+(2\bD_u+\bA^{\mathsf T}\bA)\,\dd t.
\]
Write its increment from the identity as $\bX$. Since $\bX^2=4\bD_\alpha^2\,\dd t+o(\dd t)$,
\[
  \tfrac12\log(\bI+\bX)
  =\tfrac12\bX-\tfrac14\bX^2+o(\dd t),
\]
which gives the first line of \cref{eq:hencky}. Replacing $\bF^{\mathsf T}\bF$ by $\bF\bF^{\mathsf T}$ gives the second line. Taking the trace and using $\tr(\bA^{\mathsf T}\bA)-2\tr(\bD_\alpha^2)=-\tr(\bA^2)$ gives \cref{eq:logJ}. The same trace follows from the spatial expression.

\section{Source-covariation calculation}
\label{app:source}

For constant density and divergence-free $\balpha$, the martingale momentum source is $\bQ_1=\rho(\balpha\cdot\nabla)\bu$. In components,
\[
 \rho^{-1}\partial_j[(Q_1)_i\alpha_j]
 =\partial_j(\alpha_k\partial_k u_i\,\alpha_j)
 =\partial_j(a_{jk}\partial_k u_i),
\]
because $\partial_j\alpha_j=0$. This is the covariance flux required by the full SRTT drift. If the source--transport quadratic variation is omitted, the remaining drift is not the full source-consistent balance.

\section{Green--Kubo proof}
\label{app:gk-proof}

For stationary \cref{eq:operator-ou}, the covariance at positive lag is $\E[\bq(t)\otimes\bq(0)]=e^{-Gt}\bSigma_q$. Integrating over positive and negative lags gives
\[
  \int_{-\infty}^{\infty}\E[\bq(t)\otimes\bq(0)]\,\dd t
  =G^{-1}\bSigma_q+\bSigma_qG^{-1}.
\]
Multiplying the Lyapunov equation $G\bSigma_q+\bSigma_qG=GAG$ on the left and right by $G^{-1}$ gives \cref{eq:gk}.

\section{Passive-wall covariance bound}
\label{app:passive}

For a scalar passive wall, the zero-frequency response amplitude is proportional to $(Z_f+Z_\Gamma)^{-1}$. With $B_\Gamma^2=2\vartheta Z_\Gamma$, the displacement covariance is \cref{eq:passive-wall}. Differentiation gives
\[
  \frac{\dd}{\dd Z_\Gamma}A_\Gamma^{\rm pass}
  =\frac{2\vartheta(Z_f-Z_\Gamma)}{(Z_f+Z_\Gamma)^3}.
\]
The unique maximum occurs at impedance matching $Z_\Gamma=Z_f$, where $A_\Gamma^{\rm pass}=\vartheta/(2Z_f)$. Below this maximum, the quadratic equation for $Z_\Gamma$ has two positive roots.

\section{Observer-covariance calculation}
\label{app:observer}

The spatial gradient transforms according to $\nabla^*=\bm Q\nabla$. Hence
\[
  \nabla^*\bu^*=\bm Q(\nabla\bu)\bm Q^{\mathsf T}+\bm\Omega,
  \qquad
  \nabla^*\balpha^*=\bm Q(\nabla\balpha)\bm Q^{\mathsf T}.
\]
Taking symmetric and skew parts gives
\[
  \bD_u^*=\bm Q\bD_u\bm Q^{\mathsf T},
  \qquad
  \bD_\alpha^*=\bm Q\bD_\alpha\bm Q^{\mathsf T},
  \qquad
  \bW_\alpha^*=\bm Q\bW_\alpha\bm Q^{\mathsf T}.
\]
Every quadratic term in \cref{eq:hencky,eq:qlog} is therefore conjugated by $\bm Q$. The observer spin $\bm\Omega$ is absent because it is skew and enters only the resolved rotation. This proves \cref{prop:objectivity} coefficient by coefficient.

\section{Logarithmic midpoint identity}
\label{app:log-midpoint}

Write $\bA=\bD_\alpha+\bW_\alpha$. The material and spatial quadratic drifts are
\[
\begin{aligned}
 \bQ_H&=\tfrac12\bA^{\mathsf T}\bA-\bD_\alpha^2,\\
 \bQ_h&=\tfrac12\bA\bA^{\mathsf T}-\bD_\alpha^2.
\end{aligned}
\]
Expanding the products gives
\[
\begin{aligned}
 \bQ_H&=-\tfrac12(\bD_\alpha^2+\bW_\alpha^2)
        +\tfrac12(\bD_\alpha\bW_\alpha-\bW_\alpha\bD_\alpha),\\
 \bQ_h&=-\tfrac12(\bD_\alpha^2+\bW_\alpha^2)
        -\tfrac12(\bD_\alpha\bW_\alpha-\bW_\alpha\bD_\alpha).
\end{aligned}
\]
Their arithmetic mean is \cref{eq:qlog}. The gap between the raw spatial and logarithmic-corotational routes is one half of the rotation--stretch commutator.

\section{Index-one block reduction}
\label{app:dae}

Let the time-discrete residual be $\bm F(\bm y,\bm z)=0$, where $\bm y$ contains differential velocity-like coefficients and $\bm z$ contains pressure, martingale compatibility and constraint coefficients. Linearisation gives
\[
  \bA_{\Dt}\delta\bm y+\bB\delta\bm z=-\bm F_d,
  \qquad
  \bC\delta\bm y+\bD\delta\bm z=-\bm F_a.
\]
A minimum-norm algebraic correction is
\[
  \delta\bm z=\bD^\dagger(-\bm F_a-\bC\delta\bm y)+\delta\bm z_{\rm gauge},
\]
where $\delta\bm z_{\rm gauge}$ belongs to the exact null space. Substitution gives the Schur equation
\[
  \bS_{\Dt}\delta\bm y
  =-\bm F_d+\bB\bD^\dagger\bm F_a.
\]
The DAE is index one when the algebraic equations determine $\bm z$ from $\bm y$ up to known gauge directions and the rank remains locally constant. Diagonal penalties are not equivalent: they alter the null space and can move descriptor eigenvalues.

\section{Resolved-energy derivation}
\label{app:energy-proof}

For constant density, let the resolved velocity satisfy
\[
  \rho\,\dd\bu=\bQ_0\,\dd t+\bQ_1\,\dd W_t
\]
in material coordinates. It\^o's formula gives
\[
  \rho\,\dd\frac{|\bu|^2}{2}
  =\bu\cdot\bQ_0\,\dd t+\bu\cdot\bQ_1\,\dd W_t
   +\frac{|\bQ_1|^2}{2\rho}\,\dd t.
\]
Integrating over the material volume yields \cref{eq:resolved-energy}. If $\bQ_r=\Div\bT_r+\rho\bm b_r$, then
\[
  \int_\Omega \bu\cdot\Div\bT_r\,\dd V
  =\int_{\partial\Omega}\bu\cdot(\bT_r\bn)\,\dd a
   -\int_\Omega \bT_r\inner\nabla\bu\,\dd V.
\]
For symmetric $\bT_r$, the last term is $-\bT_r\inner\bD_u$. The same calculation holds in the drift and martingale channels. This derivation is why the energy equation must start from momentum rather than from $\dd\cT\inner\dd\bm h$.

\section{Trace-class threshold}
\label{app:trace-proof}

For the isotropic spectrum $A_s\sim(\ell^{-2}+|\bm k|^2)^{-s}$ and relaxation $G_r\sim(\ell_g^{-2}+|\bm k|^2)^r$, the noise covariance in \cref{eq:operator-ou} scales as
\[
  G_rA_sG_r\sim |\bm k|^{4r-2s}
\]
at high wavenumber. The number of modes in a shell of radius $k$ is $O(k^{d-1})$. The injection trace behaves as
\[
  \int_1^\infty k^{d-1+4r-2s}\,\dd k,
\]
which converges exactly when $d-1+4r-2s<-1$, or $s>2r+d/2$. The weaker condition $s>d/2$ makes $A_s$ trace class, but it does not by itself make the forcing trace finite.

\section{State-dependent fast--slow correction}
\label{app:homogenisation}

Consider the formal fast--slow system
\[
\begin{aligned}
 \dd\zeta_i^\tau&=F_i(\bm\zeta^\tau)\,\dd t
 +C_{ij}(\bm\zeta^\tau)q_j^\tau\,\dd t,\\
 \dd\bq^\tau&=-\tau_c^{-1}\cS(\bm\zeta^\tau)\bq^\tau\,\dd t
 +\tau_c^{-1}B(\bm\zeta^\tau)\,\dd\bm W_t.
\end{aligned}
\]
At frozen $\bm\zeta$, the fast covariance is $J_f/(2\tau_c)$ up to the chosen normalisation. Solving the cell problem gives the slow response $\cD=C\cS^{-1}$. Differentiating the response along the slow stochastic motion and averaging the fast quadratic product produces
\[
  \cN_i=\partial_{\zeta_\ell}\cD_{ij}\,C_{\ell k}(J_f)_{kj}.
\]
The formula is stated at the formal homogenisation level. A full infinite-dimensional theorem requires regularity, spectral-gap and tightness assumptions beyond the scope of this preprint.

\section{Dynamic-boundary energy ledger}
\label{app:boundary-energy}

Take the bulk unresolved drift before reaction to be $\bm f_q$. From \cref{eq:wall-port}, the fluid kinetic-energy increment contains $\langle\bq,L\,\dd\bm\Lambda_\Gamma\rangle=(T\bq)\cdot\dd\bm\Lambda_\Gamma$. The wall kinetic-energy increment contains $-\bg\cdot\dd\bm\Lambda_\Gamma$. The sum is zero on the constraint $T\bq=\bg$.

The remaining wall terms are impedance dissipation, noise injection and martingale work. If $M_\Gamma$ is constant and positive,
\[
\begin{aligned}
 \dd\frac12\bg^{\mathsf T}M_\Gamma\bg
 ={}&-\bg^{\mathsf T}Z_\Gamma\bg\,\dd t
 +\frac12\Tr(M_\Gamma^{-1}B_\Gamma B_\Gamma^{\mathsf T})\,\dd t\\
 &+\bg^{\mathsf T}B_\Gamma\,\dd\bm W_t
 -\bg\cdot\dd\bm\Lambda_\Gamma.
\end{aligned}
\]
This identity distinguishes the physical traction port from the coordinate term generated by a time-dependent lifting.

\section{Field-level boundary DAE}
\label{app:field-dae}

A Chebyshev--Fourier realisation uses a homogeneous wall-normal basis and two lifting functions. Let $M_q$ and $K_q$ be the unresolved mass and relaxation matrices, and let $E$ be the discrete trace. A representative semi-discrete system is
\begin{equation}
\begin{aligned}
 \dd\bw&=(-R\bw+C\bq)\,\dd t,\\
 M_q\dd\bq&=(-C^{\mathsf T}\bw-K_q\bq)\,\dd t+E^{\mathsf T}\dd\bm\Lambda_\Gamma,\\
 M_\Gamma\dd\bg&=-Z_\Gamma\bg\,\dd t+B_\Gamma\dd\bm W_t-\dd\bm\Lambda_\Gamma,\\
 E\bq&=\bg.
\end{aligned}
\label{eq:field-boundary-dae}
\end{equation}
The adjoint pair $C,-C^{\mathsf T}$ cancels inter-sector work, while the reaction cancels through the algebraic trace constraint. This finite-dimensional system is a structural verification of the boundary port. It is not a substitute for the full nonlinear channel-flow residual.

\section{Formal multi-channel extension}
\label{app:multichannel}

For independent Brownian motions $W_t^r$ and volatility gradients $\bA_r=\nabla\balpha_r$, the local It\^o quadratic terms sum over channels. For example,
\[
  \dd\bm h
  =\sum_r\bD_r\,\dd W_t^r
  +\left[\bD_u+\frac12\sum_r\bA_r\bA_r^{\mathsf T}-\sum_r\bD_r^2\right]\dd t.
\]
The logarithmic Jacobian becomes
\[
  \dd\log J
  =\sum_r(\nabla\cdot\balpha_r)\,\dd W_t^r
  +\left[\nabla\cdot\bu-\frac12\sum_r\tr(\bA_r^2)\right]\dd t.
\]
These are formal local It\^o identities. A multi-channel logarithmic-corotation theorem may also depend on L\'evy-area data, so the one-channel work-conjugacy result is not extended here without further rough-path structure.

%% file: references.bib
@article{Memin2014,
  author={M\'emin, Etienne}, title={Fluid flow dynamics under location uncertainty},
  journal={Geophysical \& Astrophysical Fluid Dynamics}, year={2014}, volume={108}, number={2}, pages={119--146},
  doi={10.1080/03091929.2013.836190}}

@article{Holm2015,
  author={Holm, Darryl D.}, title={Variational principles for stochastic fluid dynamics},
  journal={Proceedings of the Royal Society A}, year={2015}, volume={471}, number={2176}, pages={20140963},
  doi={10.1098/rspa.2014.0963}}

@article{ResseguierEtAl2017a,
  author={Resseguier, Valentin and M\'emin, Etienne and Chapron, Bertrand},
  title={Geophysical flows under location uncertainty, Part I: Random transport and general models},
  journal={Geophysical \& Astrophysical Fluid Dynamics}, year={2017}, volume={111}, number={3}, pages={149--176},
  doi={10.1080/03091929.2017.1310210}}

@article{DrivasHolm2020,
  author={Drivas, Theodore D. and Holm, Darryl D.},
  title={Circulation and energy theorem preserving stochastic fluids},
  journal={Proceedings of the Royal Society of Edinburgh Section A: Mathematics}, year={2020}, volume={150}, number={6}, pages={2776--2814},
  doi={10.1017/prm.2019.43}}

@article{StreetCrisan2021,
  author={Street, Oliver D. and Crisan, Dan}, title={Semi-martingale driven variational principles},
  journal={Proceedings of the Royal Society A}, year={2021}, volume={477}, number={2247}, pages={20200957},
  doi={10.1098/rspa.2020.0957}}

@incollection{TissotEtAl2024,
  author={Tissot, Gilles and M\'emin, Etienne and Jamet, Quentin},
  title={Stochastic compressible Navier--Stokes equations under location uncertainty},
  booktitle={Stochastic Transport in Upper Ocean Dynamics II}, publisher={Springer}, address={Cham}, year={2024},
  doi={10.1007/978-3-031-40094-0_14}}

@book{TruesdellNoll2004,
  author={Truesdell, Clifford and Noll, Walter}, title={The Non-Linear Field Theories of Mechanics}, edition={3},
  publisher={Springer}, year={2004}, doi={10.1007/978-3-662-10388-3}}

@article{XiaoEtAl1997,
  author={Xiao, Heng and Bruhns, Otto T. and Meyers, A.},
  title={Logarithmic strain, logarithmic spin and logarithmic rate},
  journal={Acta Mechanica}, year={1997}, volume={124}, number={1--4}, pages={89--105},
  doi={10.1007/BF01213020}}

@article{Norris2008,
  author={Norris, Andrew N.}, title={Eulerian conjugate stress and strain},
  journal={Journal of Mechanics of Materials and Structures}, year={2008}, volume={3}, number={2}, pages={243--260},
  doi={10.2140/jomms.2008.3.243}}

@article{Haller2016,
  author={Haller, George}, title={Dynamic rotation and stretch tensors from a dynamic polar decomposition},
  journal={Journal of the Mechanics and Physics of Solids}, year={2016}, volume={86}, pages={70--93},
  doi={10.1016/j.jmps.2015.10.002}}

@article{Murdoch2003,
  author={Murdoch, A. I.},
  title={Objectivity in classical continuum physics: a rationale for discarding the principle of invariance under superposed rigid body motions in favour of purely objective considerations},
  journal={Continuum Mechanics and Thermodynamics}, year={2003}, volume={15}, number={3}, pages={309--320},
  doi={10.1007/s00161-003-0121-9}}

@article{GrmelaOttinger1997,
  author={Grmela, Miroslav and \"Ottinger, Hans Christian},
  title={Dynamics and thermodynamics of complex fluids. I. Development of a general formalism},
  journal={Physical Review E}, year={1997}, volume={56}, number={6}, pages={6620--6632},
  doi={10.1103/PhysRevE.56.6620}}

@article{Ottinger1998,
  author={\"Ottinger, Hans Christian},
  title={General projection operator formalism for the dynamics and thermodynamics of complex fluids},
  journal={Physical Review E}, year={1998}, volume={57}, number={2}, pages={1416--1420},
  doi={10.1103/PhysRevE.57.1416}}

@article{SerranoEspanol2001,
  author={Serrano, Mar and Espa\~nol, Pep},
  title={Thermodynamically consistent mesoscopic fluid particle model},
  journal={Physical Review E}, year={2001}, volume={64}, number={4}, pages={046115},
  doi={10.1103/PhysRevE.64.046115}}

@article{Lidstrom2012,
  author={Lidstr\"om, Per},
  title={On the principle of virtual power in continuum mechanics},
  journal={Mathematics and Mechanics of Solids}, year={2012}, volume={17}, number={5}, pages={516--540},
  doi={10.1177/1081286511426916}}

@article{BruhnsXiaoMeyers2002,
  author={Bruhns, Otto T. and Xiao, Heng and Meyers, A.},
  title={New results for the spin of the Eulerian triad and the logarithmic spin and rate},
  journal={Acta Mechanica}, year={2002}, volume={155}, pages={95--109},
  doi={10.1007/BF01170842}}

@article{KellyMelbourne2016,
  author={Kelly, David and Melbourne, Ian},
  title={Smooth approximation of stochastic differential equations},
  journal={The Annals of Probability}, year={2016}, volume={44}, number={1}, pages={479--520},
  doi={10.1214/14-AOP979}}

@article{DiamantakisWoodfield2025,
  author={Diamantakis, Theo and Woodfield, James},
  title={L\'evy Areas, Wong--Zakai Anomalies in Diffusive Limits of Deterministic Lagrangian Multitime Dynamics},
  journal={SIAM Journal on Applied Dynamical Systems}, year={2025}, volume={24}, number={1}, pages={836--893},
  doi={10.1137/24M1637271}}

@article{TissotEtAl2026,
  author={Tissot, Gilles and M\'emin, Etienne and Jamet, Quentin},
  title={Stochastic compressible Navier--Stokes equations under location uncertainty and their approximations for ocean modelling},
  journal={arXiv preprint arXiv:2309.12077v3}, year={2026},
  eprint={2309.12077}, archivePrefix={arXiv}, primaryClass={physics.flu-dyn},
  note={Revised 21 May 2026}}

@book{Pope2000,
  author={Pope, Stephen B.},
  title={Turbulent Flows},
  publisher={Cambridge University Press},
  year={2000}}

@article{Smagorinsky1963,
  author={Smagorinsky, Joseph},
  title={General circulation experiments with the primitive equations. I. The basic experiment},
  journal={Monthly Weather Review}, year={1963}, volume={91}, number={3}, pages={99--164},
  doi={10.1175/1520-0493(1963)091<0099:GCEWTP>2.3.CO;2}}

@article{GermanoEtAl1991,
  author={Germano, Massimo and Piomelli, Ugo and Moin, Parviz and Cabot, William H.},
  title={A dynamic subgrid-scale eddy viscosity model},
  journal={Physics of Fluids A}, year={1991}, volume={3}, number={7}, pages={1760--1765},
  doi={10.1063/1.857955}}

@article{MeneveauKatz2000,
  author={Meneveau, Charles and Katz, Joseph},
  title={Scale-invariance and turbulence models for large-eddy simulation},
  journal={Annual Review of Fluid Mechanics}, year={2000}, volume={32}, pages={1--32},
  doi={10.1146/annurev.fluid.32.1.1}}

@article{HottovyEtAl2015,
  author={Hottovy, Scott and McDaniel, Austin and Volpe, Giovanni and Wehr, Jan},
  title={The Smoluchowski--Kramers Limit of Stochastic Differential Equations with Arbitrary State-Dependent Friction},
  journal={Communications in Mathematical Physics},
  year={2015}, volume={336}, number={3}, pages={1259--1283},
  doi={10.1007/s00220-014-2233-4}}

@article{BirrellEtAl2017,
  author={Birrell, Jeremiah and Hottovy, Scott and Volpe, Giovanni and Wehr, Jan},
  title={Small Mass Limit of a Langevin Equation on a Manifold},
  journal={Annales Henri Poincar\'e},
  year={2017}, volume={18}, number={2}, pages={707--755},
  doi={10.1007/s00023-016-0508-3}}
